\documentclass[a4paper]{article}
\usepackage[leqno]{amsmath}
\usepackage{amssymb} %Ez kell $\mathbb R$-hez.
\usepackage{amsthm}  %Ez meg a proof kornyezethez.
\newcommand*{\hop}{\bigskip\\}
\renewcommand*{\int}{\intop\limits}
\newcommand{\fej}[1]{\section{\!\!\!\!\!\!.\ #1}}
\newcommand{\alfej}[1]{\subsection{\!\!\!#1}}
\newcommand{\nemfej}[1]{\renewcommand\thesection{}
\section{\!\!\!\!\!\!#1}\renewcommand\thesection{\arabic{section}}}

\def\Di{\text D}
\def\ttau{\boldsymbol\tau}
\def\M{\boldsymbol M}
\def\di{\text d}
\def\I{\boldsymbol I}
\def\pt{\partial}
\def\a{\alpha}
\def\b{\beta}
\def\g{\gamma}
\def\bs{\boldsymbol}
\def\E{\bs E}
\def\g{\omega}
\def\vp{\varphi}
\def\Lf{\mathfrak L}
\def\Hh{\bs H}
\def\L{\bs L}
\def\D{\bs D}
\def\Rb{\mathbb R}
\def\s{\sigma}
\newcommand\e[1]{\text{e}^{#1}}

\begin{document}
\title{Lagrange-formalism of point-masses}
\author{M\'arton Bal\'azs\footnote{Institute of Mathematics, BUTE}\\
P\'eter V\'an\footnote{Department of Chemical Physics, BUTE}}
\maketitle

\begin{abstract}
We prove by symmetry properties that the Lagrangian of a free
point-mass is a quadratic function of the speed in the
non-relativistic case, and that the action of the free point-mass
between two spacetime points is the proper time passed in the
relativistic case. These well known facts are proved in a
mathematically rigorous way with a frame independent treatment
based on spacetime models introduced by Matolcsi. The arguments
show that these results are not obvious at all, some common
beliefs can be refuted by explicit counterexamples. In our
treatment the similarity of non-relativistic and relativistic
cases is apparent.
\end{abstract}

\nemfej{Introduction}

Spacetime symmetries play a fundamental role in physical theories.
We can distinguish between two points of view regarding their
application.
If governing equations are known, symmetries provide conserved
quantities (symmetry charges) of the theory, conditions on
their conservation and integrals of the governing equations.
When the dynamics is given by variational principles the
symmetries are exploited by Noether theorems. If governing
equations are unknown, then symmetries are the basic tools to
construct a proper dynamics (based on any kind of formulation)
of the theory. The starting point is always to determine the
Lagrangian of free systems (point-mass or field). A free system
is invariant under all symmetries of the corresponding
spacetime and it is completely determined by this property.

In classical mechanics, where the governing equation -- the Newton
equation -- is well known one can exploit symmetries by Noether
theorems (see e.g. \cite{ArnAta89b,Gol80b,AbrMar78b}). On the
other hand, in quantum field theories symmetries are common tools
to construct appropriate Lagrangians (see. e.g. \cite{Ram90b}). A
similar -- exceptional -- approach in classical mechanics can be
found in   the books of Landau and Lifsic where the method is
applied for finding the Lagrangian of a point-mass in
non-relativistic dynamics \cite{LanLif76b} and partially also in
relativistic dynamics \cite{LanLif78b}.

That derivation (and all similar subsequent derivations in field
theories) are unsatisfactory from several points of view. Although
the Lagrangians are coordinate free, they are observer and
therefore frame dependent. Moreover,  it is not clear
whether the action, the Lagrangian, the resulted Euler-Lagrange
equations or the solutions of the Euler-Lagrange equations are the
transformed objects that should be invariant under the symmetries.
Different papers and textbooks give different answers, moreover
these cases are frequently mixed (and contradictory) in the very
same treatment.

We shall see that if the Lagrangian or the action itself were
required to be invariant for all spacetime symmetries (Noether or
Poincar\'e transformations), then we should get a constant
function. Thus it is essential that instead of the Lagrangians,
the invariant quantities should be the equivalence classes of the
Lagrangians (two Lagrangians differing in a full time derivative
are considered equivalent). Somewhere in the literature this
problem is obliterated by referring to the fact that two
Lagrangians in the same equivalence class can be transformed into
each other with a canonical transformation \cite[p26]{Ram90b}.
That is true, but it has nothing to do by invariance because we
cannot transform the variables anyhow when investigating
invariance under variable transformations (spacetime symmetries).

The situation is even more confused in the (special) relativistic
case where the observer dependence of the Lagrangian is more
apparent (there is no frame independent time). There symmetry
considerations are essentially ignored \cite{LanLif78b} with the
argument that the only invariant scalar is proper time. If
symmetry is taken into account, the invariance of the solutions
of Euler-Lagrange equations is required \cite{Run66b} resulting
in a set of basically different Lagrange functions of a single free
point-mass.

In this paper only the dynamics of {\em free} point-masses is
considered, therefore we require invariance under the full
spacetime symmetry group. Our treatment can give surprising
results constructing dynamics of interacting systems, when only
partial spacetime symmetries are required. First of all in our
frame independent spacetime models it is apparent that some of the
partial spacetime symmetries are observer dependent (there is no
time translation, time inversion etc. without a frame), therefore
they cannot be considered as fundamental.
Independently of the above mentioned property of partial
symmetries, our more exact study gives some real surprises:
we shall show e.g. by an explicit example that spacetime translation
symmetry alone does not imply momentum conservation.

In the present paper we investigate the precise restrictions of
the possible forms of the Lagrangian of a free point-mass, implied
by symmetry assumptions. We consider two Lagrangians equivalent if
they result in the same Euler-Lagrange equations i.e. differ in a
full time derivative. We use arguments both in relativistic and
non-relativistic case which are correct from a mathematical point
of view as well. We show that the problem in these two cases can
in fact be handled in a very similar way. We use the spacetime
model constructed without observers and reference frames,
described in Matolcsi \cite{Mat84b,Mat93b}, as a general and
sophisticated tool for our observations. The formalism is based on
a clever exploitation of the affine structure of non-relativistic
and special relativistic spacetimes giving a method that is
similar to the traditional tensorial one. A differential geometric
treatment like \cite{SalVit00a} would be unnecessarily technical
and not very well fitted to this problem. Another advantage of our
treatment is to avoid misunderstandings based on the the well
known problematic formulation of general covariance in
non-relativistic and special relativistic theories \cite{Fri83b}.
Spacetime symmetries are frequently explained by the equivalence
of inertial reference frames; the free point-mass is said to be
invariant under translations, rotations and velocity
transformations {\sl of the reference frames}. Why just the
movements of the {\em free} point-mass would be invariant under
such changes? Working without reference frames of course excludes
similar problems: spacetime transformations are really acting on
the spacetime position and velocity of the particle.

\fej{Point-mass in non-relativistic spacetime}

\alfej{The non-relativistic spacetime}

Here we recall some basic structures of the frame independent
formalism described in Matolcsi \cite{Mat93b}. All the vector
spaces in question are real. The non-relativistic spacetime is
$(M, I, \tau, \D, {\bs b})$, where
\begin{itemize}
\item[- $M$] is {\sl spacetime}, an oriented four-dimensional
affine space over the vector space ${\M}$, i.e.\ the difference
of two points in $M$ is an element of $\M$. By orientation we
mean a fixed ordering of basis vectors of ${\M}$.

\item[- $I$] is {\sl time}, an oriented one-dimensional affine
space over ${\I}$. The latter is a vector space, {\sl the
measure line of time intervals}. By its orientation, we have
positive and negative time intervals. Between any two moments
of $I$, the time interval is an element of ${\I}$.

\item[- $\tau$] $:\,M\to I$ is an affine surjection, making
correspondence between a point of $M$ and its absolute time in
$I$. ${\ttau}\,:\,{\M}\to {\I}$ is the linear surjection under
$\tau$, connecting each vector to its time interval. These two
functions are the {\sl time evaluation functions}.

\item[- $\bs D$] is an oriented one-dimensional vector space,
{\sl the measure line of distances}.

\item[- ${\bs b}$] $:\,\bs E\times\bs E\to\bs D\otimes\bs D$ is
a positive definite symmetric bilinear mapping, the {\sl
Euclidean structure}, where $\bs E:\,=\ker\ttau$ is the {\sl
subspace of the spacelike vectors}.
\end{itemize}

We obtain the usual coordinate-description by vectorizing $M$ with
an origin in $M$ and a basis in $\M$. Coordinates are denoted by
roman letters, and can have values $0,\,1,\,2,\,3$. These are
written in superscript, while coordinates of the dual space $\M^*$
are put in subscripts. We use Einstein-convention for summing
indices in superscripts and subscripts denoted by the same letter.
Vectors in $\E$ have first coordinate zero, we denote the other
three indices by Greek letters.

The history of a point-mass is described by a {\sl world line},
a connected curve with timelike tangents (tangents not in
$\E$). Such a world line can be given as the range of a {\sl
world line function}, a continuously differentiable function
$r\,:\,I\to M$ defined on an interval, with
$\tau(r(t))=t$ for any $t\in\text{Dom}\,r$. Using a
construction similar to the one of tensor products, one can
define the four-dimensional tensor quotient space
$\frac{\M}{\I}$. The derivative of a world line function is
$\dot r\,:\,I\to\frac{\M}{\I}$ with $\ttau(\dot r(t))=1$. Hence
its values are elements of
\[
V(1):\,=\left\{\bs u\in\frac{\M}{\I}\,:\,\ttau(\bs u)=1\right\},
\]
the set of {\sl absolute velocity values}. $V(1)$ is an affine
space over $\frac{\E}{\I}$, and $\ddot
r(t)\in\frac{\E}{\I\otimes\I}$ if $r$ is twice differentiable
at $t\in I$.

We need the set of mass values, which we define as follows. $\I$
measures time lengths, it contains the {\sl second}, while $\D$
measures distances, containing {\sl meter}. For simplicity we
choose $\hbar:\,=1$, hence mass values (e.g.\
$1\,\frac{\text{second}}{\text{meter}^2}$) are in
$\frac{\I}{\D\otimes\D}$, the measure line of mass values
\cite{SalVit00a}.

\alfej{Galilean and Noether transformations}

Noether transformations are the automorphisms of the spacetime.
These transformations keep the structure of spacetime, and can be
described as follows. The {\sl proper Galilean group} is
\begin{multline*}
\mathcal G:\,=\bigl\{\L\in\text{Lin}(\M,\M)\,:\,\L\
\text{preserves orientation},\,\ttau\cdot\L=\ttau,\\
\L|_{\E}^{\phantom{\E}+}\cdot\L|_{\E}=\text{id}_{\E}\bigr\}
\end{multline*}
acting on $\M$, and rotating spacelike vectors. The {\sl proper
Noether group} (inhomogeneous Galilean group) is
\[
\mathcal N:\,=\bigl\{L\,:\,M\to M\,:\, L\text{\ is affine, and
the underlying}\ \L\ \text{is an element of}\ \mathcal
G\bigr\}.
\]
The word proper refers to the fact that time or space inversion
is not contained in these groups.

Spacetime translations i.e. transformations of the form $x\mapsto
x+\bs a$ with a given $\bs a\in \M$ are Noether transformations,
whose underlying linear operator is the identity of $\M$.
$\mathcal G$ and $\mathcal N$ are a six dimensional and a ten
dimensional Lie group, respectively. They have the Lie algebras
\begin{eqnarray*}
\text{La}(\mathcal
G)&=&\bigl\{\Hh\in\text{Lin}(\M,\M)\,:
\,\ttau\cdot\Hh=0,\,\Hh|_{\E}^+=-\Hh|_{\E}\bigr\}\text{\
\ and}\\
\text{La}(\mathcal N)&=&\bigl\{H\,:\,M\to\M\,: \,H\text{\
affine, and the underlying}\ \Hh\ \text{is in La}(\mathcal
G)\bigr\},
\end{eqnarray*}
respectively. By its definition, an $\Hh\in\text{La}(\mathcal G)$
is in fact an $\M\to\E$ linear map. The Lie algebra of the
subgroup of spacetime translations consists of elements $H$ for
which $\Hh=0$. Such an affine map is constant, i.e. there is an
$\bs h\in\M$ such that $Hx=\bs h$ for all $x\in Mw$.

Every Noether transformation $L$ in a neighborhood of the unit
element $\text{id}_M$ has the form
\begin{equation}
L=\e{s H}:\,=I+\sum_{n=1}^{\infty}\frac{(s\Hh)^{n-1} \cdot s
H}{n!}\label{eq:miazl}
\end{equation}
for some $s\in\Rb$ and $H\in\text{La}(\mathcal N)$. The underlying
Galilean transformation is
\[
\L=\e{s\Hh}:\,=\sum_{n=0}^\infty\frac{(s\Hh)^n}{n!}.
\]

\alfej{Variational principle for point-masses \\ in
non-re\-la\-ti\-vis\-tic spacetime}\label{sc:var}

According to variational principles of mechanics, the point-mass
moves along a world line from a spacetime point $x_0$ to another
one $x_1$, for which the ``variation'' of an ``action function''
is zero, possibly caused by having extremum or stationary value of
the action function at this world line. In case of hamiltonian
variational principles the initial and final spacetime points are
fixed (the duration and the initial and final space points are not
varied). We formulate this as follows.

Given a Lagrangian depending continuously on spacetime points
and speed values, and mapping to the one-dimensional vector
space $\frac{\Rb}{\bs I}$:
\[
\Lf\,:\,M\times V(1)\to\frac{\Rb}{\bs I},
\]
the action on an $r$ world line function is
\[
S(r):\,=\int_{t_0}^{t_1}\Lf\bigl(r(t),\,\dot r(t)\bigr)dt
\]
with $t_0:\,=\tau(x_0)$ and $t_1:\,=\tau(x_1)$.

In order to use analysis arguments, we define differentiability
of the function $r\mapsto S(r)$. Taking a norm $|\ \ |_{\M}$ on
$\M$ and a norm $|\ \ |_{\frac{\M}{\I}}$ on $\frac{\M}{\I}$ (any
two norms on a finite dimensional vector space are equivalent).
We introduce the vector space
\begin{equation}
\bs V{:=}\{\bs r:[t_0,t_1]\to\E\mid \bs r\ \text{is
continuously differentiable}, \bs r(t_0){=}\bs r(t_1){=}0
\}\label{eq:vvvt}
\end{equation}
endowed with the norm
\[
\|\bs r\|:\,=\max_{t\in [t_0, t_1]}\left(|\bs
r(t)|_{\M}+|\dot{\bs r}(t)|_{\frac{\M}{\I}}\right).
\]
Then
\[
V:=\{r:[t_0,t_1]\to M\mid r\ \text{is a world line function},\
r(t_0)=x_0,\ r(t_1)=x_1\}
\]
is an affine space over $\bs V$. Hence differentiability of
$S\,:\,V\to\Rb$ is well-defined.

If $S$ is differentiable, the world line function realized by
the point-mass is selected by
\[
\Di S(r)=0\in\text{Lin}(\bs V,\,\Rb),
\]
i.e.\ the derivative of $S$ having value zero. This corresponds
to the ``action having variation zero''. It is well known, that if
the Lagrangian $\Lf$ is twice continuously differentiable, then
$S$ is differentiable, and in this case, $\Di S(r)=0$ is
equivalent to twice continuous differentiability of $r$
satisfying the Euler-Lagrange equation
\[
\Di_1\Lf(r(t),\,\dot r(t))-\frac{\di}{\di t}\Di_2\Lf(r(t),\,\dot r(t))=0.
\]
Here $\Di_1$ stands for the partial derivative according to the
first variable in $M$, and $\Di_2$ for the derivative according
to the second variable in $V(1)$. From now on, we denote
elements of $M$ by $x$, and elements of $V(1)$ by $u$,
therefore we write
\[
\frac{\pt\Lf(x,u)}{\pt
x}\in\text{Lin}(\M,\,\Rb/\I),\qquad\frac{\pt\Lf(x,u)}{\pt
u}\in\text{Lin}(\E/\I,\Rb/\I)
\]
for the partial derivatives, respectively.

By the construction of this variational principle, it is clear
that adding a ``full time-derivative'' to $\Lf$ only means
adding a constant to $S$, hence leaving $\Di S$ invariant
together with the world line realized. We give precise meanings
of these notions.

A function $\mathfrak f\,:\,M\times V(1)\to\frac{\Rb}{\I}$ is
called a {\sl full time-derivative} if there exists a
$\phi\,:\,M\to\Rb$ continuously differentiable function, such
that $\mathfrak f(x,u)=\Di\phi(x)u$ for all $x\in M$ and $u\in
V(1)$. For any world line function $r,\ \mathfrak f\circ(r,\dot
r)=(\phi\circ r)\dot{}$ holds in this case, hence the action
corresponding to $\Lf$ and to $\Lf+\mathfrak f$ only differ by
a constant. We say that $\Lf$ and $\Lf'$ are {\sl equivalent},
if $\Lf'-\Lf$ is a full time-derivative. This relation
determines equivalence classes on the set of Lagrangians.

\alfej{Symmetries and the Lagrangian}

A motion of a physical system happens at the same way before and
after a transformation, if the solution of the Euler-Lagrange
equation is not affected by the transformation. We assume that not
only these solutions, but the derivative $\Di S$ characterizes as
well the physics of the system. Hence we say that a transformation
is a symmetry of the system, if it leaves the derivative $\Di S$
invariant, i.e.\ it only turns the Lagrangian into an equivalent
one. This is a rather restrictive standpoint, there are several
examples for variational principles where only the equivalence of
the solutions is required e.g. in non-equilibrium thermodynamics
where the governing equations cannot be derived from a variational
principle, therefore the usual variational prescriptions are
shaken up \cite{VanMus95a}, but also in mechanics there are
attempts to find physical consequences of that fact investigating
the so called s-equivalent systems (\cite{CisLop01a} and the
references therein). Now we exclude transformations leaving the
solution invariant, but multiplying $\Lf$ by a constant for
example.

Let $F\,:\,M\to M$ be a continuously differentiable map, for
which $\ttau\cdot\Di F(x)u\neq0$ for any $u\in V(1)$. We say
that $F$ is a {\sl symmetry} of the Lagrangian $\Lf$, if there
exists a  full time-derivative $\mathfrak f_F$ such that
\begin{equation}
\Lf\left(Fx,\,\frac{\Di F(x)u}{\ttau\cdot\Di
F(x)u}\right)\ttau\cdot\Di F(x)u=\Lf(x,\,u)+\mathfrak
f_F(x,\,u)\label{eq:symm}
\end{equation}
for any $x\in M,\ u\in V(1)$. The definition considers that the
transformation can change the (absolute) time and assures that a
transformed world line function remain a world line function after
the transformation (e.g. the second variable of $\Lf$ is an
element of $V(1)$ and the integration should change to leave $DS$
invariant after a reparameterization of the time scale). The
definition is more transparent in case of symmetries that do not
rescale the absolute time ($\ttau\cdot\Di F(x)u = 1$). The above
definition is valid uniformly in special relativistic and
non-relativistic considerations, too. Non-relativistic space-time
symmetries do not rescale the (absolute) time. Especially, a
proper Noether transformation is a symmetry of $\Lf$, if and only
if
\[
\Lf(Lx,\,\L u)=\Lf(x,\,u)+\mathfrak f_L(x,\,u),
\]
since $\Di L(x)=\L$, and $\ttau\cdot\L=\ttau,\ \ttau u=1$ by
definition of the proper Galilean group and of $V(1)$.

\alfej{Lagrangian of a free point-mass}

If a point-mass is free, i.e.\ it is not influenced by any effect,
then we ``feel'' that the translated, rotated etc.\ form of its
trajectory is also a possible trajectory for it. To be more
precise, we could say that by applying a spacetime-automorphism on
a trajectory selected by the variational principle, we obtain
another trajectory satisfying that principle. It is still not a
precise statement, since it is not clear how to understand a
``free'' point-mass, ``not influenced by any effect''. We reverse
the situation, and accept this concept as a definition.

We define a point-mass characterized by $\Lf$ to be a {\sl free
point-mass}, if each proper Noether transformation is a symmetry
of $\Lf$. Hence $\Lf$ is the Lagrangian of a free point-mass if
and only if for any $L\in\mathcal N$ there exists a
$\phi_L\,:\,M\to\Rb$, for which
\[
\Lf(Lx,\,\bs Lu)-\Lf(x,\,u)=\Di\phi_L(x)u.
\]

We introduce the notation $\widehat\phi(L,\,x):=\phi_L(x)$, and we
assume that $\widehat\phi\,:\,\mathcal N\times M\to\Rb$ is smooth
enough. Although $\mathcal N\times M$ is not an affine space thus
$\widehat\phi$ is defined on a manifold, we only consider
one-parameter subgroups of $\mathcal N$, hence we can use the
usual differentiability notions. We take the elements in the
neighborhood of $I:\,=\text{id}_M$ in the form \eqref{eq:miazl},
and we differentiate by the parameter $s$. Since
\[
\frac{\di\e{sH}}{\di s}\biggr|_{s=0}=H\text{\ \ and \ \ \
}\frac{\di\e{s\Hh}}{\di s}\biggr|_{s=0}=\Hh,
\]
we obtain
\begin{equation}
\frac{\pt\Lf(x,\,u)}{\pt x}\cdot H(x)+\frac{\pt\Lf(x,\,u)}{\pt
u}\cdot\Hh\cdot u=
\frac{\pt^2\widehat\phi(L,\,x)}{\pt L\,\pt x}\biggr|_{L=I}\cdot(H,\,u)=
\,:\frac{\pt\g(H,\,x)}{\pt x}\cdot
u\label{eq:elso}
\end{equation}
by letting $s\to0$.

In order to compare our frame independent formulae with those of
usual treatments, we write the coordinated forms of our
expressions. We coordinate $\M$ by an appropriate basis, we
vectorial $M$ by the map $x\mapsto x-o$ with a fixed point $o\in
M$, and we consider the vector-coordinates of these vectors. Then
$H(x)=\bs H(x-o)+\bs h$, where $\bs h:\,=H(o)\in\M$. If $\Hh$ has
coordinates ${H^i}_j$, then $H(x)$ has coordinates
${H^i}_jx^j+h^i$. Let us remark here that sometimes one think on
coordinates as a convenient tool for expressing tensorial
calculations without the corresponding reference frames (see the
concept of "abstract indexes" of Wald \cite{Wal84b}). However, a
formulation of general covariance (observer independence) with an
observer dependent notation easily can lead to misinterpretation
because a frame independent equation can lead to a formula
containing observer dependent quantities in a particular reference
frame (especially in non-relativistic spacetime, see the debate on
the covariance of the kinetic theory e.g.
\cite{Mul72a,Mur83a,LebBou88a,MatGru96a,Mus98a}). Therefore,
although there is no convenient notation to book the different
transposes of higher order tensors without indexes, it is
important to formulate the results of the calculations in our
frame independent notation, too.

$V(1)$ is an affine subspace in $\frac{\M}{\I}$, the four
coordinates of its elements are not independent, i.e.\ the zeroth
coordinates are $1$ in this space. Hence the derivatives by
elements of $V(1)$ only contain indices $1,\,2,\,3$. We will
distinguish these possibilities in the notation. Greek letters
are in $\{1,2,3\}$ and Latin indexes in $\{0,1,2,3\}$. Double
indexes denote summation. Therefore, the coordinated form of
\eqref{eq:elso} is
\[
\frac{\pt \Lf}{\pt x^i}({H^i}_jx^j+h^i)+\frac{\pt\Lf}{\pt
u^\a}{H^\a}_ju^j=\frac{\pt\g}{\pt x^i}u^i,
\]
without writing the arguments of our functions.

First we consider the special case of spacetime-translations. Then
$\Hh=0$, hence $H(x)=\bs h\in\M$ is the same constant for each
$x\in M$. Therefore, we can write \eqref{eq:elso} in the form
\[
\frac{\pt\Lf(x,\,u)}{\pt x}\cdot\bs h=\frac{\pt\g(\bs h,\,x)}{\pt
x}\cdot u \ \ ,\qquad\frac{\pt\Lf}{\pt x^i}h^i=\frac{\pt\g}{\pt
x^i}u^i.
\]
The left hand-side is linear in $\bs h$, the right hand-side is
linear in $u$. Hence the other sides also have these properties.
Thus it follows that there are functions $l, f:M\to\M^*$ for which
\[
\frac{\pt\g(\bs h,\,x)}{\pt x}={\bs h}\cdot D f(x)\
\ ,\qquad\frac{\pt\g}{\pt x^i}=\frac{\pt f_j}{\pt x^i}h^j,
\]
and
\[
\frac{\pt\Lf(x,\,u)}{\pt x}=u \cdot D l(x)\  \
,\qquad\frac{\pt\Lf}{\pt x^i}=\frac{\pt l_j}{\pt x^i}u^j.
\]
Therefore,
\begin{equation}
Dl(x) = (Df)^*(x)\ \ ,\qquad \frac{\pt l_i}{\pt x^j}=\frac{\pt
f_j}{\pt x^i}. \label{eq:masod}\end{equation}

Assuming twice differentiability of $\Lf$, we differentiate
\eqref{eq:masod} by $x^k$. Changing the order of
differentiation and applying \eqref{eq:masod} again we obtain by
Young's theorem
\[
\frac{\pt^2 f_j}{\pt x^k \pt x^i} =
\frac{\pt^2 l_i}{\pt x^k \pt x^j} =
\frac{\pt^2 l_i}{\pt x^j \pt x^k} =
\frac{\pt^2 f_k}{\pt x^j \pt x^i}.
\]
As a result, we can get that
\[
D(Df - (Df)^*)(x)=0 \ , \qquad
\frac{\pt}{\pt x^i}\biggl(\frac{\pt f_j}{\pt x^k}-\frac{\pt
f_k}{\pt x^j}\biggr)=0.
\]

Hence introducing the antisymmetric linear map $\bs
C:\M\to\M^*$ with components $C_{jk}$, we obtain
\begin{equation}
\bs C:\,=-\Di\land f:\,=\Di f-(\Di f)^*=\text{const.} \ ,\qquad
C_{jk}:\,=\frac{\pt f_j}{\pt x^k}-\frac{\pt f_k}{\pt
x^j}=\text{const.}
\label{eq:constC}\end{equation}

Therefore
\begin{equation}
Dl(x)= Df(x) + \bs C^* \ , \qquad
\frac{\pt l_j}{\pt x_k} = \frac{\pt f_j}{\pt x_k} + C_{kj}.
\label{eq:lf}\end{equation}

We conclude
\[
\Lf(x,\,u)=f(x)\cdot u+(x-o)\cdot\bs C\cdot u+\vp_f(u)\ ,\quad
\Lf=f(x)_ku^k+x^kC_{kj}u^j+\vp_f(u),
\]
\noindent with an arbitrary function $\vp_f\,:\,V(1)\to\Rb/\I$.
Exploiting (\ref{eq:lf}) one may write also
\[
\Lf(x,\,u)=l(x)\cdot u + \vp_l(u)\ \
,\quad\Lf=l(x)_ku^k + \vp_l(u),
\]
with an arbitrary function $\vp_l\,:\,V(1)\to\Rb/\I$.
Hence we obtained from the spacetime translation symmetry that
\begin{equation}
\Lf(x,\,u)=l(x)\cdot u+\vp(u)\ \ ,\qquad\Lf=l(x)_ku^k+\vp(u),
\label{eq:leltol}
\end{equation}
where $l$ is a function satisfying
\begin{equation}
D \land l = {\bf C} = \text{const.} \ , \qquad
\frac{\pt^2 l_j}{\pt x^k \pt x^i} =
    \frac{\pt^2 l_i}{\pt x^k \pt x^j}.
\label{eq:gs}\end{equation}

This is all we can say; we could not prove the statement of
\cite[page 13]{LanLif76b}; from spacetime translation symmetry
does not follow the independence of the Lagrangian of spacetime
variables.

Now we consider general Noether transformations, and substitute
the form \eqref{eq:leltol} of the Lagrangian into
\eqref{eq:elso} (recall that $\Hh$ is a map $\M\to\E$):
\[
\frac{\pt l(x)\cdot u}{\pt x}\cdot H(x)+l(x)\cdot\Hh\cdot
u+\frac{\di\vp(u)}{\di u}\cdot\Hh\cdot u=\frac{\pt\g(H,x)}{\pt
x}\cdot u,
\]
i.e.
\begin{equation}
\frac{\pt l_k}{\pt
x^i}u^k({H^i}_jx^j+h^i)+l_{\a}{H^{\a}}_ku^k+\frac{\pt\vp}{\pt
u^\a}{H^\a}_ku^k=\frac{\pt\g}{\pt x^k}u^k.
\label{eq:altn}\end{equation}

Differentiating this by $x^m$ leads to
\[
\frac{\pt^2l_k}{\pt x^m\pt x^i}u^k({H^i}_jx^j+h^i) +
\frac{\pt l_k}{\pt x^i}u^k{H^i}_m +
\frac{\pt l_{\a}}{\di x^m}{H^\a}_k u^k =
\frac{\pt^2\g}{\pt x^m\pt x^k}u^k.
\]
The variable $u$ is only present linearly in this equation,
hence we can omit it:
\begin{equation}
\frac{\pt^2 l_k}{\pt x^m\pt x^i}({H^i}_jx^j+h^i)+
\frac{\pt l_k}{\pt x^i}{H^i}_m +
\frac{\pt l_\a}{\di x^m}{H^\a}_k =
\frac{\pt^2\g}{\pt x^m\pt x^k}.
\label{eq:itt}\end{equation}
We transpose the equation (consider it with the indices $k$ and $m$
interchanged), and subtract it from the original form
\eqref{eq:itt}. Then the right hand-side is zero by Young's
theorem, and the first term on the left hand-side disappears
due to \eqref{eq:gs}:
\[
\frac{\pt l_k}{\pt x^i}{H^i}_m-\frac{\pt l_m}{\pt
x^i}{H^i}_k+\frac{\pt l_{\alpha}}{\pt x^m}{H^{\alpha}}_k-\frac{\pt
l_{\alpha}}{\pt x^k}{H^{\alpha}}_m=0.
\]
As we know, $\Hh$ is a linear map $\M\to\E$, hence
${H^0}_k=0$. Therefore, we can sum over indices
$\a=1,2,3$ instead of $i=0,\dots,3$ contained in the
expressions ${H^i}_k$:
\[
\frac{\pt l_k}{\pt x^{\a}}{H^{\a}}_m-\frac{\pt l_m}{\pt
x^{\a}}{H^{\a}}_k+\frac{\pt l_{\a}}{\pt x^m}{H^{\a}}_k-\frac{\pt
l_{\a}}{\pt x^k}{H^{\a}}_m=0,
\]
i.e.
\begin{equation}
C_{k\a}{H^{\a}}_m - C_{\a m}{H^{\a}}_k=0.
\label{eq:ch}\end{equation}

We know that restricting $\Hh$ to $\E$ results in an
antisymmetric map, and by the identification
$\E^*\equiv\frac{\E}{{\bs D}\otimes{\bs D}}$, we can
interchange subscripts and superscripts of spacelike vectors:
${H^\a}_\b=-{H_\b}^\a$. Thus
separating the cases $k=\b=1,2,3$ and $k=0$, we obtain
 for $m=\g=1,2,3$,
\begin{equation}
C_{\b\a}{H^\a}_\g - C_{\g\a}{H^\a}_\b=0,
\label{eq:komm}\end{equation}
\begin{equation}
C_{0\a}{H^\a}_\g + C_{\g\a}{H^\a}_0=0.
\label{eq:cs}\end{equation}

Let $\bs i:\E\to\M$ be the embedding map; its transpose $\bs
i^*:\M^*\to\E^*$ is a linear surjection. Then
$\bs C_{\E}:=\bs i^*\bs C\bs i:\E\to\E^*\equiv\dfrac{\E}{{\bs D}\otimes{\bs
D}}$ is an antisymmetric linear map. Equation \eqref{eq:komm}
tells us that this commutes with any antisymmetric linear map
$\Hh|_{\E}:\E\to\E$:
\[
\left[\bs C_{\E},\,\Hh|_{\E}\right]=0.
\]
As a consequence, $\bs C_{\E}$ commutes with all the elements of the
rotation group of $\E$, hence by Schur's lemma, it is a
multiple of $\text{id}_{\E}$, meanwhile it is antisymmetric.
This is only possible if
\[
\bs C_{\E}=0\ \ ,\qquad C_{\a\b}=0.
\]
Then, by \eqref{eq:cs}, $C_{0\a}{H^\a}_\g=0$ for each
$\Hh$, which implies
\[
C_{0\a}=0,
\]
finally
\[
\bs C=0.
\]
Then \eqref{eq:constC} implies that the derivative of $l$ is
symmetric (i.e.\ the antisymmetric derivative is zero), hence
$l$ is a derivative of some function $\phi\,:\,M\to\Rb$:
\[
l=\Di\phi\ \ ,\qquad l_k=\frac{\pt\phi}{\pt x^k}.
\]
Thus the first term in expression \eqref{eq:leltol} of the
Lagrangian is a full time-derivative, which can be omitted. We
conclude
\begin{equation}
\Lf(x,\,u)=\vp(u).\label{eq:lseb}
\end{equation}
Now it is clear that we needed not only translation invariance,
but rotation invariance as well in order to exclude spacetime
dependence of the Lagrangian. Writing \eqref{eq:elso} again, now
we get
\begin{equation}
\Di\vp(u)\cdot\Hh\cdot u=\frac{\pt\g(H,\,x)}{\pt x}\cdot u\ \
,\qquad\frac{\pt\vp}{\pt u^\a}{H^\a}_ju^j=\frac{\pt\g}{\pt
x^i}u^i.\label{eq:harmad}
\end{equation}
First we consider maps $\Hh$, for which $\Hh|_{\E}=0$. For each
of these maps, there exists a $\bs v\in\frac{\E}{\I}$, such
that $\Hh\cdot u=\bs v$ for all $u\in V(1)$. Then
\[
\Di\vp(u)\cdot\bs v=\frac{\pt\g(H,\,x)}{\pt x}\cdot u\ \
,\qquad\frac{\pt\vp}{\pt u^\a}v^\a=\frac{\pt\g}{\pt x^i}u^i.
\]
The left hand-side does not depend on $x$, and contains $\bs v$
linearly. Hence the same holds for the right hand-side as well.
Therefore there is a linear map
$\bs A:\frac{\M}{\I}\to\E^*\equiv\frac{\E}{{\bs D}\otimes{\bs D}}$
such that
\[
\Di\vp(u)=\bs A\cdot u\ \
,\qquad \frac{\pt \vp}{\pt u^\a}= A_{\a i}u^i \ .
\]
Let $\bs A_{\E}$ denote the restriction of $\bs A$ onto
$\frac{\E}{\I}$; then fixing an element $c$ of $V(1)$ and
putting  $a:=\bs A c$, we have
\[
\Di\vp(u)=\bs A_{\E}\cdot(u-c) + a\ ,\qquad
\frac{\pt \vp}{\pt u^\a} = A_{\a\b}(u^\b -c^\b) + a_\a \ .
\]
Differentiating by $u^\b$, we infer from
Young's theorem that $\bs A_{\E}$ is symmetric i.e.
$A_{\a\b}=A_{\b\a}$ and a simple calculation results in
\[
\vp(u)=\frac1{2}\bigl((\bs A_{\E}\cdot(u-c)\bigr)\cdot(u-c) +
a\cdot(u-c)\ + \text{const.}. \label{eq:uas}
\]
The last terms here are full time-derivatives, which can be omitted.
Using this form let us return to formula \eqref{eq:elso}:
\[
(\bs A_{\E}(u-c))\cdot\Hh\cdot
u=\frac{\pt\g(H,x)}{\pt x}\cdot u\ \ .
\]
The left hand side is zero at $u=c$, thus the right hand side,
too, hence the previous formula can be written as
\[
(\bs A_{\E}\cdot(u-c))\cdot\Hh\cdot(u-c)=g(H,x)\cdot (u-c) \ .
\]
The left hand side is bilinear in $u-c$, the right hand
side is linear; this is possible only if the right hand side is
zero. Note that here in fact the restriction of $\Hh$ onto
$\E$ appears; the well-known properties of the antisymmetric
maps $\Hh|_{\E}:\E\to\E$ imply that $(\bs
A_{\E}\cdot(u-c))\cdot\Hh\cdot(u-c)=0$
can hold for all $\Hh$ only if $\bs
A_{\E}(u-c)$ is parallel to $(u-c)$ i.e. there is an
$m\in\dfrac{\I}{{\bs D}\otimes{\bs D}}$ (mass value)
such that
$$\vp(u)=\frac1{2}m|u-c|^2\ \ .$$

\alfej{Reformulating the variational principle}\label{sc:trukk}

We show a method which we could have used but which would not
have had any advantage. We only introduce this method here
because we need to apply it later in the relativistic case, and
we can show similarities to that case.

A world line of a point-mass has been given as the range of a
world line function so far. This is in fact not necessary, since
we can parameterize such a curve in many different ways. Let $\bs
1$ be any fixed positive element of $\I$, and for a world line
function $r$, let $p:[\bs 0,\,\bs 1]\to\text{Ran}\,r$ be a
parameterization for the corresponding world line. By the inverse
function theorem, $\s:\,=r^{-1}\circ p:[\bs 0,\bs 1]\to[t_0,t_1]$
is a continuously differentiable bijection, and
\[
\dot p=(\dot r\circ\s)\dot\s,
\]
for which applying $\ttau$ we obtain
\[
0<\ttau\circ\dot p=\dot\s.
\]
The function $\s$ is given by $p$ (as a primitive function of
$\ttau\circ\dot p$), hence $r$ is also determined by $p$. We
have
\[
r\circ\s=p\ \ ,\quad\dot r\circ\s=\frac{\dot p}{\ttau\circ\dot p}\ \ ,
\]
and
\[
\ddot r\circ\s=\frac{\ddot p}{(\ttau\circ\dot p)^2}-\frac{\dot
p\,\ttau\circ\ddot p}{(\ttau\circ\dot p)^3}
\]
holds as well in case of twice differentiability. We see that $r$
satisfies a second-order differential equation if and only if $p$
does so.

The action according to this new parameterization can be
computed using integral transformation:
\begin{multline}
\int_{t_0}^{t_1}\Lf\bigl(r(t),\,\dot r(t)\bigr)\,\di
t=\int_{\bs 0}^{\bs 1}\Lf\bigl(r(\s(a)),\,\dot
r(\s(a))\bigr)\dot\s(a)\,\di a=\\
=\int_{\bs 0}^{\bs 1}\Lf\left(p(a),\,\frac{\dot
p(a)}{\ttau\cdot\dot p(a)}\right)\ttau\cdot\dot p(a)\,\di
a.\label{eq:paramh}
\end{multline}
According to the new parameterization, $\dot p$ is allowed to
be an arbitrary future-like vector, not only an element of
$V(1)$. Let $N^{\to}\subset\frac{\M}{\I}$ denote the set of
future-like vectors. This is an open set containing $V(1)$. By
the integral transformation above, the Lagrangian can be
considered as a function
\[
\Lf\,:\,M\times N^{\to}\to\frac{\Rb}{\I}\ \ ,\quad
(x,\,w)\mapsto\Lf(x,\,w)
\]
having the property
\[
\Lf(x,\,w)=\Lf\left(x,\,\frac{w}{\ttau\cdot w}\right)\ttau\cdot w.
\]
Hence the Lagrangian $\Lf:M\times V(1)\to\frac{\Rb}{\I}$
used so far is now extended to a function $M\times
N^{\to}\to\frac{\Rb}{\I}$  according to
\eqref{eq:paramh}; the action itself only depends on the world
lines, not on their parameterization.

Now we consider the vector space
\[
\bs V:=\{\bs p:[\bs 0,\,\bs 1]\to\M\mid\bs p\ \text{is
continuously differentiable},\ \bs p(\bs 0)=\bs p(\bs 1)=0\},
\]
endowed with the norm
\[
||\bs p||:\,=\max_{s\in[\bs 0,\,\bs 1]}(|\bs
p(s)|_{\M}+|\dot{\bs p}(s)|_{\frac{\M}{\I}})
\]
and the affine space
\[
V:=\{p:[\bs 0,\bs 1]\to M\mid p\ \text{is continuously
differentiable},\ p(\bs 0)=x_0,\,p(\bs 1)=x_1\}
\]
over $\bs V$. For the action
\[
S\,:\,V\to\Rb\ \ ,\quad p\mapsto\int_{\bs 0}^{\bs
1}\Lf(p(s),\,\dot p(s))\,\di s,
\]
we can repeat everything we said before, and its extremal points
give the extremal points of the original action function.

The original Lagrangian is a restriction of this new one to the
set $M\times V(1)$. Hence it is clear that the extended
Lagrangians $\Lf$ and $\Lf'$ differ by a full time-derivative
if and only if their restrictions do so:
$\Lf'(x,\,u)-\Lf(x,\,u)=\Di\vp(x)u$ for all $(x,u)\in M\times
V(1)$ is equivalent to $\Lf'(x,\,w)-\Lf(x,\,w)=\Di\vp(x)w$ for
all $(x,\,w)\in M\times N^{\to}$. By definition \eqref{eq:symm}
of the symmetries, it is clear that we call $F$ a symmetry of
the extended Lagrangian, if
\[
\Lf(Fx,\,\Di F(x)w)=\Lf(x,\,w)+\mathfrak f_F(x,\,w)
\qquad((x,\,w)\in M\times N^{\to}).
\]

\fej{Point-mass in relativistic spacetime}

\alfej{The relativistic spacetime}

This model is also introduced and described in details in Matolcsi
\cite{Mat93b}. The relativistic spacetime is $(M,\,\I,\,g)$, where
\begin{itemize}

\item[- $M$] is {\sl spacetime}, a four-dimensional oriented
real affine space over the vector space $\M$.

\item[- $\I$] is {\sl the measure line of time intervals}, an
oriented one-dimensional real vector space.

\item[- $g$] is an $\M\times\M\to\I\otimes\I$ arrow-oriented
{\sl Lorentz form}.

\end{itemize}
\noindent
With the use of the speed of light, physical distances can be
identified with time intervals, hence there is no need for a
new measure line besides $\I$ (we use the ``unit system''
$\hbar=c=1$).

The motion of a particle is described by a {\sl world line}, a
connected curve with timelike tangents. A world line is
naturally given as the range of a {\sl world line function}.
The latter is a continuously differentiable function
$r\,:\,\I\to M$ defined on an interval, and
\[
\dot r(t)\in V(1):\,=\left\{u\in\frac{\M}{\I}\biggm| u\cdot
u=-1, u\text{\ is future-like}\right\}.
\]
 An essential difference is, compared with the
non-relativistic case, that $V(1)$ is not an affine space here.

The time passed {\sl along a world line} can be measured as
follows. Let $x_0$ and $x_1$ be two points of a world line $C$.
We consider an arbitrary parameterization $p$ of this world
line, having for simplicity the domain $[\bs 0,\,\bs 1]$ and
values $p(\bs 0)=x_0,\ p(\bs 1)=x_1$. The time passed between
these two points on the world line $C$ is
\[
t_C(x_0,\,x_1)=\int_{\bs 0}^{\bs 1}\sqrt{|\dot p(a)\dot
p(a)|}\,\di a.
\]
Especially, if the parameterization is the world line function
$r$, then $t_C(x_0,\,x_1)=r^{-1}(x_1)-r^{-1}(x_0)$.

\alfej{Lorentz and Poincar\'e transformations}

Transformations preserving the structure of the spacetime,
i.e.\ the automorphisms of spacetime are called the {\sl proper
Poincar\'e transformations}. They can be described as follows.
\begin{multline*}
\mathcal L:\,=\bigl\{\bs L\in\text{Lin}(\M,\,\M)\,|\,\L\text{\
is orientation and arrow-orientation preserving},\\
\L^+\cdot\bs L=\text{id}_{\M}\bigr\}
\end{multline*}
is {\sl the proper Lorentz group}, and
\[
\mathcal P:\,=\bigl\{L\,:\,M\to M\,|\,L\text{\ is affine, and
the underlying}\ \bs L\in\mathcal L\bigr\}
\]
is {\sl the proper Poincar\'e group}. The word ``proper'' relies
to the absence of time or space inversion in these groups.
Spacetime translations, i.e.\ transformations of the form
$x\mapsto x+\bs a$ (with $\bs a\in\M$) are Poincar\'e
transformations whose underlying linear operator is the identity
of $\M$.

$\mathcal L$ and $\mathcal P$ are a six dimensional and a ten
dimensional Lie group. They have the Lie algebras
\[
\renewcommand\arraystretch{1.2}
\begin{array}{c}
\text{La}(\mathcal
L)=\bigl\{\Hh\in\text{Lin}(\M,\,\M)\,|\,\Hh^+=-\Hh\bigr\}\text{\
\ and}\\
\text{La}(\mathcal P)=\bigl\{H\,:\,M\to\M\,|\,H\text{\ is
affine, and the underlying\ }\Hh\in\text{La}(\mathcal
L)\bigr\},
\end{array}
\]
respectively.

The Lie algebra of the subgroup of spacetime translations
consists of elements $H$ for which $\Hh=0$.
Such an affine map is a constant, i.e.\ there is an $\bs
h\in\M$, such that $Hx=\bs h$ for all $x\in M$.

Every Poincar\'e transformation $L$ in the neighborhood of the unit
element $\text{id}_M$ has the form
\begin{equation}
L=\e{sH}\,:\,=I+\sum_{n=1}^{\infty}\frac{(s\Hh)^{n-1}\cdot
H}{n!}\label{eq:rmiazl}
\end{equation}
for some $s\in\Rb$ and $H\in\text{La}(\mathcal L)$. The
underlying Lorentz transformation is
\[
\L=\e{s\Hh}=\sum_{n=0}^\infty\frac{(s\Hh)^n}{n!}.
\]

\alfej{The variational principle in relativistic spacetime}

In a naive way, we would say that given a continuous function
$\Lf\,:\,M\times V(1)\to\frac{\Rb}{\I}$, a point-mass
propagates from a spacetime point $x_0$ to another one $x_1$
along a world line, on which the action
\begin{equation}
S\,:\,r\mapsto
S(r)=\int^{r^{-1}(x_1)}_{r^{-1}(x_0)}\Lf\bigl(r(t),\,\dot
r(t)\bigr)\,\di t\label{eq:naivs}
\end{equation}
is extremal. But this leads to a very serious problem. Along
different world lines connecting $x_0$ and $x_1$, different
amount of time passes, hence the domains of such world line
functions are not common. The set of such world line functions
can not be made an affine space directly, we cannot use the
methods of differential calculus. Would we step over these
problems somehow, there would be another problem with the
Euler-Lagrange equation: $V(1)$ is not an affine space, hence
differentiating by the speed variable would only be possible
using the theory of manifolds.

We help this problem by the method shown in section
\ref{sc:trukk}. World lines will be parameterized by an
arbitrary parameter having domain $[\bs0,\,\bs1]\subset\I$. For
a world line function $r$, let
$p\,:\,[\bs0,\,\bs1]\to\text{Ran}r$ be a parameterization of
the corresponding world line. Then $\s:\,=r^{-1}\circ
p\,:\,[\bs0,\,\bs1]\to[t_0,\,t_1]$ is a continuously
differentiable bijection ($t_0=r^{-1}(x_0)$\ and\
$t_1=r^{-1}(x_1)$), and
\[
\dot p=(\dot r\circ\s)\dot\s\ \ ,\qquad 0<|\dot
p|:\,=\sqrt{-\dot p\cdot\dot p}=\dot\s.
\]
Therefore, $\s$ is determined by $p$ (as the primitive function
of $|\dot p|$), $r$ can also be restored from $p$, and
\[
r\circ\s=p\ \ ,\quad\dot r\circ\s=\frac{\dot p}{|\dot p|};
\]
in case of twice differentiability we have
\[
\ddot r\circ\s=\frac{\ddot p}{|\dot p|^2}+\frac{\dot p\,(\dot
p\cdot\ddot p)}{|\dot p|^4}.
\]
We see that $r$ obeys a second-order differential equation if and
only if $p$ does so. We allow any parameterization as described
above, hence its derivative can be any future-like vector, not
only an element of $V(1)$. Let $N^{\to}\subset\frac{\M}{\I}$ be
the set of the future-like vectors. This is an open set,
containing $V(1)$.

We introduce the vector space
\[
\bs V:=\{\bs p\,:\,[\bs 0,\,\bs1]\to\M\mid\bs p\text{\ is
continuously differentiable},\ \bs p(\bs0)=\bs p(\bs1)=0\}
\]
endowed with the norm
\[
||\bs p||:\,=\max_{a\in[\bs0,\,\bs1]}(|\bs
p(a)|_{\M}+|\dot{\bs p}(a)|_{\frac{\M}{\I}})
\]
and the affine space
\[
V:=\{p\,:\,[\bs0,\,\bs1]\to M\mid p\text{\ is continuously
differentiable},\ p(\bs0)=x_0,\,p(\bs1)=x_1\}
\]
over $\bs V$.

We are given a twice differentiable Lagrangian
\[
\Lf\,:\,M\times N^{\to}\to\frac{\Rb}{\I}\ \
,\qquad(x,\,w)\mapsto\Lf(x,\,w),
\]
satisfying
\begin{equation}
\Lf(x,\,w)=\Lf\left(x,\,\frac{w}{|w|}\right)|w|.
\label{eq:rskala}\end{equation}
We assume that the particle moves
in such a way that the parameterization $p$ of its world line
between points $x_0$ and $x_1$ in spacetime makes the derivative
$\Di S(p)$ of the action
\[
S\,:\,V\to\Rb\ \ ,\quad
p\mapsto\int_{\bs0}^{\bs1}\Lf(p(s),\,\dot p(s))\,\di s
\]
zero. Equation \eqref{eq:rskala} is trivial in the physically
relevant cases $w\in V(1)$, since $|w|=1$ for such speed values.
Hence \eqref{eq:rskala} only gives properties of the extension of
the Lagrangian from $V(1)$ to $N^{\to}$, it does not mean any
extra physical condition on the Lagrangian. By \eqref{eq:rskala}
and by replacing the integral variable, it is easy to show that
the action is independent of the parameterization $p$ of the world
line.

We can repeat all our arguments we had in section \ref{sc:var};
the parameterization $p$ of the world line realized satisfies the
appropriate Euler-Lagrange equation. The time scale invariance
property of special relativistic point-mass dynamics is a well
known problem of mechanics. The Lagrangians satisfying
(\ref{eq:rskala}) are sometimes called "homogeneous" and this is
why one cannot formulate Hamilton equations and canonical
transformations in special relativistic mechanics without any
further ado i. e. the formal Hamiltonian defined analogously to
the non-relativistic case is identically zero (see e.g.
\cite{Run66b}). From our treatment one can see clearly that the
scale invariance property cannot be avoided, contrary to the
opinion of Goldstein \cite[p329]{Gol80b}: "... it is not a
sacrosanct physical law that the action integral in Hamilton's
principle must have the same value wether expressed in terms of
$t$ of in terms of $\theta$ ... All that is required is that $L$
be a world scalar that leads to the correct equations of motion."
If we gave a clear meaning to the expression "leads" and without
knowing it we looked for a "correct" equation of motion, then we
should exclude every Lagrangian where the action would depend on
the parameterization of the world lines.

The definition of symmetries (\ref{eq:symm}) and the scale
invariance property treated in section \ref{sc:trukk} was a
somewhat artificial construction, because in non-relativistic
spacetime the time is absolute. However, the striking similarity
of the non-relativistic and relativistic concepts deserves the
attention.

\alfej{Symmetries and the Lagrangian}

We call a {\sl full time-derivative} a function $\mathfrak
f\,:\,M\times N^{\to}\to\frac{\Rb}{\I}$, if there exists a
function $\phi\,:\,M\to\Rb$ such that $\mathfrak
f(x,\,w)=\Di\phi(x)w$. Two Lagrangians determine the same
variational principle (are equivalent), if and only if their
difference is a full time-derivative, i.e.\ the derivative $\Di
S$ connected to them agree.

We say that a continuously differentiable map $F\,:\,M\to M$, for
which $\Di F(x)w\in N^{\to}$ for each $w\in N^{\to}$, is a {\sl
symmetry} of the (homogeneous) Lagrangian, if there is a full
time-derivative $\mathfrak f_F$, such that
\[
\Lf(Fx,\,\Di F(x)w)=\Lf(x,\,w)+\mathfrak
f_F(x,\,w)\qquad((x,\,w)\in M\times N^{\to}).
\]
Especially, a  proper Poincar\'e transformation $L$ is a
symmetry of $\Lf$, if
\[
\Lf(Lx,\,\L w)=\Lf(x,\,w)+\mathfrak f_L(x,\,w).
\]

\alfej{Lagrangian of the free point-mass}

Similarly to the non-relativistic case, we call the point-mass
{\sl free}, if each Poincar\'e transformation is a symmetry of the
corresponding Lagrangian $\Lf$. Hence $\Lf$ is the Lagrangian of a
free point-mass, if and only if for all $L\in\mathcal P$ there is
a function $\phi_L\,:\,M\to\Rb$ such that
\[
\Lf(Lx,\,\bs Lw)-\Lf(x,\,w)=\Di\phi_L(x)w.
\]
We introduce the notation $\widehat\phi(L,\,x):\,=\phi_L(x)$, and
we assume that this function $\mathcal P\times M\to\Rb$ is smooth
enough. Although $\widehat\phi$ is defined on the manifold
$\mathcal P\times M$, we only consider one-parameter subgroups of
$\mathcal P$, hence we can use our usual differentiability
notions.

We consider transformations in a neighborhood of
$I:\,=\text{id}_M$ in the form \eqref{eq:rmiazl}. We obtain
\begin{multline}
\frac{\pt\Lf(x,\,w)}{\pt x}\cdot H(x)+\frac{\pt\Lf(x,\,w)}{\pt
w}\cdot\Hh\cdot w=\frac{\pt^2\widehat\phi(L,\,x)}{\pt L\pt
x}\biggr|_{L=I}\cdot(H,\,w)=\,:\frac{\pt\g(H,\,x)}{\pt x}\cdot
w\label{eq:relso}
\end{multline}
the same way as in the non-relativistic case. We write the
coordinated form as well:
\[
\frac{\pt\Lf}{\pt x^i}({H^i}_jx^j+h^i)+\frac{\pt\Lf}{\pt
w^i}{H^i}_jw^j=\frac{\pt\g}{\pt x^i}w^i.
\]
As in the non-relativistic case, we find
\begin{equation}
\Lf(x,\,w)=l(x)\cdot w+\vp(w)\ \
,\qquad\Lf=l(x)_kw^k+\vp(w),\label{eq:rlf}
\end{equation}
with an arbitrary function $\vp\,:\,V(1)\to\Rb/\I$ and with
$l\,:\,M\to\M^*$ for which
\[
\bs C:\,=\Di\land l:\,=(\Di l)^*-\Di l=\text{const.}\ \ ,\qquad
C_{ki}:\,=\frac{\pt l_i}{\di x^k}-\frac{\pt l_k}{\pt
x^i}=\text{const.}
\]

By considering general Poincar\'e transformations and
substituting the form \eqref{eq:rlf} to \eqref{eq:relso}, we
conclude
\[
\frac{\pt l(x)\cdot w}{\pt x}\cdot H(x)+l(x)\cdot\Hh\cdot
w+\frac{\di\vp(w)}{\di w}\cdot\Hh\cdot w=\frac{\pt\omega
(H,x)}{\pt x}\cdot w,
\]
i.e.\
\[
\frac{\pt l_k}{\pt
x^i}w^k({H^i}_jx^j+h^i)+l_i{H^i}_kw^k+\frac{\pt\vp}{\pt
w^i}{H^i}_kw^k=\frac{\pt\g}{\pt x^k}w^k.
\]
Differentiating the latter by $x^m$ leads to
\[
\frac{\pt^2l_k}{\pt x^m\pt x^i}w^k({H^i}_jx^j+h^i)+\frac{\pt
l_k}{\pt x^i}w^k{H^i}_m+\frac{\pt l_i}{\di
x^m}{H^i}_kw^k=\frac{\pt^2\g}{\pt x^m\pt x^k}w^k.
\]
By the method shown in the non-relativistic case, it follows
that
\[
C_{ki}{H^i}_m-C_{mi}{H^i}_k=0.
\]

$\Hh$ is antisymmetric and, by the identification
$\M^*\equiv\frac{\M}{{\I}\otimes{\I}}$, subscripts and superscripts
can be interchanged, hence ${H^i}_k=-{H^k}_i=-{H_k}^i$. Using
also antisymmetry of $\bs C$,
\[
C_{ki}{H^i}_m-{H_k}^iC_{im}=0,
\]
i.e.
\[
[\bs C,\,\Hh]=0.
\]
for all antisymmetric $\Hh$. As a consequence,
$\bs C$ commutes with all proper Lorentz transformations, hence
$\bs C$ is a multiple of $\text{id}_{\M}$ by Schur's lemma. On
the other hand, it is also antisymmetric, thus
\[
\bs C=0\ \ ,\qquad C_{ik}=0.
\]
Based on this, as in the non-relativistic case, we conclude
that the Lagrangian has the form
\[
\Lf(x,\,w)=\vp(w)
\]
by omitting a full time-derivative. According to \eqref{eq:rskala}
we have
\[
\vp\left(\frac{w}{|w|}\right)|w|=\vp(w).
\]
Choosing an arbitrary number $\lambda>0$ and using the previous
equality, we obtain
\[
\vp(\lambda w)=\vp\biggl(\frac{\lambda w}{|\lambda
w|}\biggr)|\lambda
w|=\lambda\vp\biggl(\frac{w}{|w|}\biggr)|w|=\lambda\vp(w);
\]
differentiating with respect to $\lambda$ and substituting
$\lambda=1$, we get
\begin{equation}
\Di\vp(w)\cdot w=\vp(w)\ \ ,\qquad\frac{\pt\vp}{\pt
w^k}w^k=\vp.\label{eq:lam}
\end{equation}

\noindent Since the Lagrangian does not depend on spacetime
points, considering also the equivalence of the Lagrangians by a
full time derivative, its symmetry is formulated as follows: for
all Lorentz transformations $\L$ there exists an $\a(\L)\in \M^*$
such that
\begin{equation}
\vp(\L w)=\vp(w)+\a(\L)\cdot w +const..
\label{eq:vpsy}\end{equation}
We differentiate this equation with respect to $w$:
\[
\Di\vp(\L w)\L=\Di\vp(w)+\a(\L)\ \ ,\qquad\frac{\pt\vp}{\pt
w^i}(\L w){L^i}_k=\frac{\pt\vp}{\pt w^k}(w)+\a(\L)_k.
\]

Considering a Lorentz transformation for which $\L w=w$ holds, we
have $\a(\L)=0$ by \eqref{eq:vpsy}, hence for all such
transformations $\Di\vp(w)\L=\Di\vp(w)$, thus
$\L^*\Di\vp(w)=\Di\vp(w)$. By $\L^*\equiv\L^{-1}$, this means
\[
\Di\vp(w)=\L\Di\vp(w)\quad\text{if}\quad\L w=w.
\]
Since $\{\L\mid\L w=w\}$ is just the rotation group of the
three-dimensional subspace orthogonal to $w$ (i.e.\ it is a
``little group" of Wigner), the equation above can only hold in
case $\Di\vp(w)$ is parallel to $w$. This means that there
exists a function $\beta\,:\,\frac{\M}{\I}\to\Rb$ such that
\[
\Di\vp(w)=\beta(w)w\ \ ,\qquad\frac{\pt\vp}{\pt w^k}=\beta\ w_k,
\]
hence multiplying \eqref{eq:lam} with $w$, we obtain
\[
\vp\,w_i=\frac{\pt\vp}{\pt
w^k}w^k\,w_i=\beta\,w_k\,w^k\,w_i=\frac{\pt\vp}{\pt
w^i}w_k\,w^k,
\]
thus the differential equation
\[
\Di\vp(w)=\frac{\vp(w)\,w}{w\cdot w}\ \
,\qquad\frac{\pt\vp}{\pt w^i}=\frac{\vp\,w_i}{w^kw_k}
\]
holds for $\vp$. This partial differential equation can be
handled with methods of ordinary differential equations. For
the zeroth variable for example, by fixing all the three other
variables, with notations $x:\,=w^0=-w_0,\
a^2:\,=\sum\limits_{i=1}^3w^iw_i$, we have the ordinary
differential equation
\[
\frac{\di\psi}{\di x}=\frac{-\psi x}{-x^2+a^2}=\frac{\psi
x}{x^2-a^2},
\]
or for the first variable, with fixed other variables and with
notations $y:\,=w^1=w_1,\
b^2:\,=-(w^0w_0+w^2w_2+w^3w_3)=-w\cdot w+w^1w_1>0$,
\[
\frac{\di\psi}{\di y}=\frac{\psi y}{y^2-b^2}.
\]
Then it is easy to see that the solution of \eqref{eq:lam} is
\[
\vp(w)=m\,|w|,
\]
with a constant $m\in\frac{\Rb}{\I}$ (mass value).

\alfej{Discussion}

We obtained the Lagrangian
\[
\Lf(x,\,u)=\frac1{2}m|u-c|^2
\]
of a non-relativistic free point-mass (with a mass value $m$ and
an absolute velocity $c$) by considering Noether transformations
(members of the inhomogeneous Galilean group) of form $\e{sH}$
%(in a neighborhood of the identity)
but it is a simple fact that every Noether transformation is a
symmetry of the Euler-Lagrange equations based on the above
Lagrangian.

We remark that different $c$-s correspond to different but
equivalent Lagrangians, while different $m$-s result in
inequivalent Lagrangians. The Lagrangian given by $c$ is the
kinetic energy of the point-mass, relative to the inertial
observer having absolute velocity $c$.

The Lagrangian does not depend on spacetime points, hence is
invariant for spacetime translations. On the other hand, it is not
invariant for all Noether transformations: if $L$ is a
Noether transformation whose underlying linear map is the
special Galilean transformation $\L_{\bs v}$ with speed $\bs
v$, then
\[
\Lf(Lx,\,\L u)=\frac1{2}m|(u+\bs v)-c|^2=\frac1{2}m|u-(c-\bs v)|^2,
\]
i.e.\ the Lagrangian turns into the (equivalent) Lagrangian
given by  $c-\bs v\in V(1)$.

As a consequence, we see that if we required that the Lagrangian
itself be invariant for all Noether transformations, then we
should get a constant function. This fact shows well, why our
symmetry definition is preferable among all others. Evidently, we
cannot require the invariance of the action, because it can be
different for the very same Euler-Lagrange equations, a full time
derivative gives an additional constant (in field theories it is
excluded by appropriate boundary conditions). We cannot require
the invariance of the solutions, because the gained freedom is too
large \cite{VanNyi99a}. That we really require is the invariance
of the Euler-Lagrange equations that appears as an invariance of
an equivalence class of the Lagrangians.

We emphasize that spacetime homogeneity (invariance for spacetime
translations) alone does not imply that the Lagrangian does not
contain explicit spacetime dependence, contrary to usual
statements \cite{LanLif76b}: spacelike rotations, too, are
necessary to deduce this result.

We give a  simple counterexample. Choosing an ``origin" $o\in M$
and an antisymmetric linear map $\bs B\,:\,\M\to\M^*$, we define
the Lagrangian
\[
\Lf(x,\,u):\,=(x-o)\cdot\bs B\cdot u + \vp(u)
\]
according to \eqref{eq:leltol}. For $\bs B\ne0$, the first term
is not a full time-derivative, hence $\Lf$ and all Lagrangians
in its equivalence class depend explicitly on spacetime points.
On the other hand, a spacetime translation with any $\bs
a\in\M$ is a symmetry of this Lagrangian:
\[
\Lf(x+\bs a,\,u)-\Lf(x,\,u)= \bs a\cdot \bs B\cdot u
\]
and the right hand-side is a full time-derivative (of the
function $x\mapsto \bs a\cdot\bs B\cdot(x-o))$. The statement
that spacetime homogeneity implies independence of $\Lf$ of
spacetime variables is hence not true. This Lagrangian is
symmetric to spacetime translations, and contains an essential
spacetime dependence. The Euler-Lagrange equation has the form
\[
2\bs B\cdot \dot r - \Di\vp(\dot r)\cdot\ddot r=0,
\]
which is clearly invariant under spacetime translations but
momentum is not conserved.

We obtained the Lagrangian
\[
\Lf(x,\,w)=m\,|w|.
\]
of a relativistic free point-mass (with a mass value $m$) by
considering Poincar\'e transformations of form $\e{sH}$
%(in a neighborhood of the identity)
but it is a simple fact that every Poincar\'e transformation is a
symmetry of the above Lagrangian.

Lagrangians with different $m$-s are inequivalent. The Lagrangian
itself and not only its equivalence class is invariant to all
Poincar\'e transformations.  There are no other possible choices
with our assumptions. Goldstein \cite{Gol80b} admits $L=m f(|w|)$
with a two times differentiable monotonous $f: \Rb \rightarrow
\frac{\Rb}{\I}$ as a Lagrangian of a free point-mass. However, he
requires the symmetry of the solutions of Euler-Lagrange equations
in a restricted sense and admitting this possibility one can get
far more general Lagrangians (see \cite{VanNyi99a} on the
possibilities).

As in the non-relativistic case, spacetime homogeneity (invariance
for spacetime translations) alone does not imply that the
Lagrangian does not contain explicit spacetime dependence. The
above counterexample for spacetime dependence and non-conservation
of momentum can be repeated word by word.

The relativistic action depends only on the parameterized world
line, not on the pa\-ra\-me\-te\-ri\-za\-ti\-on itself. We
conclude that the action is in fact $m$ times the proper time
passed along the world line which is usually stated as
``obvious''. The Lagrangian is constant on the absolute velocity
values, i.e.\ on the set $M\times V(1)$: $\Lf(x,\,u)=m$ if $u\in
V(1)$. According to \eqref{eq:naivs}, the particle moves between
two timelike-separated points by minimizing its proper time
passed. This also shows problems with definition \eqref{eq:naivs}:
a constant Lagrangian leads to a trivial Euler-Lagrange equation,
which holds for any motion. We do not have hence Euler-Lagrange
equation for world line functions. In \cite{LanLif78b} for
example, variational principle is written for motions
parameterized by the time of an observer.

For the parameterization $p$ of a world line of a free
point-mass, we have the Euler-Lagrange equation
\[
\left(\frac{\dot p}{|\dot p|}\right)^{\bullet}=0\ \
,\quad\text{i.e.}\quad\frac{\dot p}{|\dot p|}=\text{\
constant},
\]
hence the particle has constant velocity, the corresponding
world line is a straight line.

Let us mention here, that the parameterization invariance of the
relativistic Lagrangian leads to the famous difficulties of
relativistic dynamics and connected to degeneracy and
non-covariant property of the traditional (pseudo) energy-momentum
in general relativity as well. In general relativity one can find
examples to the extension of the Lagrangian (locally from $V(1)$
to $N^{\to}$ with our notation) \cite{WalZou99a}, but a different
possibility is to work with Dirac's formalism based on constrained
variations \cite{Dir49a,Fad82a}.

Our treatment shows clearly that variational principles and
symmetries in both the non-relativistic and the relativistic case
can be handled in a very similar way; most of our arguments are
identical almost word by word. We have got the well-known results
in a rigorous way. The spacetime models and the exact formulation
of the problem helped us to see the proper reasons of these
well-known results.

\section{Acknowledgement}

The authors are very much indebted to T. Matolcsi for his
stimulating and constructive critical remarks.

\noindent
\renewcommand\arraystretch{0.5}

\small\begin{tabular}{l}
M\'arton Bal\'azs\\
Institute of Mathematics,\\
Budapest University of Technology and Economics\\
1111 Egry J\'ozsef u.\ 1, H \'ep., V. em. \\
Budapest, Hungary\\
\texttt{balazs@math.bme.hu}
\end{tabular}
\hop

\small\begin{tabular}{l}
P\'eter V\'an\\
Department of Chemical Physics,\\
Budapest University of Technology and Economics\\
1521 Budafoki \'ut.\ 8\\
Budapest, Hungary\\
\texttt{vpet@phyndi.fke.bme.hu}
\end{tabular}

\end{document}